\documentclass
{aa}
\usepackage{graphicx}
\newcommand{\ha}{H$\alpha$}
\newcommand{\kms}{km s$^{-1}~$}
\begin{document}
   \title{Stellar dynamics of blue compact galaxies\thanks{Based on observations collected at the European Southern Observatory, Paranal, Chile, under observing programme 71.B-0602.} }
\subtitle{I. Decoupled star--gas kinematics in ESO\,400-G43}

%   \subtitle{I. Overviewing the $\kappa$-mechanism}

   \author{G{\"o}ran {\"O}stlin\inst{1}
          \and
      Robert J. Cumming\inst{1}
          \and
      Philippe Amram\inst{4}
          \and
      Nils Bergvall\inst{2}
      	\and
      Daniel Kunth\inst{5}
      	\and
      Isabel M{\'a}rquez\inst{3} 
    	\and
      Josefa Masegosa\inst{3}
     	\and
      Erik Zackrisson\inst{2}}

   \offprints{G{\"o}ran {\"O}stlin \email{ostlin@astro.su.se}}

   \institute{Stockholm Observatory, AlbaNova University Center, SE-106 91 Stockholm, Sweden.              
          \and
    Uppsala Astronomical Observatory, Box 515, SE-751 20
    Uppsala, Sweden.
    \and
    Instituto Astrofisica de Andalucia (CSIC), Camino Bajo de Huetor 24,
	E-18008 Granada, Spain
    \and
	Observatoire Astronomique Marseille-Provence \& Laboratoire
	d'Astrophysique de Marseille, 2 place Le Verrier, 13248 Marseille 
	Cedex 4, France
    \and
	Institute d'Astrophysique de Paris, 98bis boulevard Arago, 75014 Paris, France
}

   \date{Received ; accepted }

   \abstract{In this {\it letter} 
we present the first results from a program aimed at measuring
the stellar kinematics of blue compact galaxies by observing
the near-infrared Calcium triplet.
%In this paper w
We show the first results for ESO\,400-G43 based on deep
VLT/FORS2 spectroscopy. The instabilities found in the central
gaseous velocity field are not seen in the solid body stellar
rotation curve, indicating that stars and gas are kinematically
decoupled in this galaxy. Even if this galaxy has a perturbed
gaseous velocity field, the stellar velocity dispersion on average
agrees well with that derived from the nebular lines. 
%This suggests
%that the latter may be used to probe the gravitational potentials
%of high redshift galaxies, too faint to be studied from their
%stellar absorption lines.

}
   \titlerunning{Stellar dynamics of BCGs: ESO\,400-G43}
   \authorrunning{{\"O}stlin et al.}
   \maketitle
%
%________________________________________________________________

\section{Introduction}

There has been strong interest in Blue Compact Galaxies (BCGs) ever
since the pioneering study of Searle \& Sargent (1972). The hypothesis
that BCGs represent genuinely young galaxies, presently forming
their first generation of stars, has now been disproved for all but a
few extreme cases (see, e.g., Kunth \& {\"O}stlin 2000 for a
review). However, the evolutionary history of BCGs, their connection
to other kinds of galaxies, and the mechanism that triggers the active
star formation are still elusive.

Most studies of the nature of BCGs have been based on surface photometry
(e.g. Papaderos et al. \cite{papaderos}) or spectroscopic analysis of  
nebular abundances (e.g. Izotov and Thuan \cite{it}), and relatively little 
is known about their dynamics, though H{\sc i} velocity fields have
been obtained for some galaxies (e.g. van Zee et al. 1998).
The stellar component has been much less analysed than the gaseous
one because the spectra are dominated by hot stars and ionised gas
emission, diluting stellar absorption features in the optical
domain and making kinematical and chemical analysis difficult.

In {\"O}stlin et al. (\cite{ostlin99}, \cite{ostlin01}) we investigated 
the velocity fields of the ionised gas in BCGs, using the CIGALE  
Fabry-Perot interferometer (Amram et al. \cite{amram91}) and targeting 
the \ha \ line. These results show that many luminous BCGs
(LBCGs) 
have peculiar \ha \ velocity fields and irregular morphology, suggesting
that their starbursts have been triggered by dwarf galaxy mergers.
Many LBCGs also rotate too slowly to support the stellar
mass inferred from deep optical and near-IR photometry ({\"O}stlin et al.
\cite{ostlin01}). This mass discrepancy can be resolved if it is assumed
that the width of the \ha \ line  traces virial motions (see Melnick et al. 
\cite{melnick87}),
indicating that these
systems may be supported by velocity dispersion, i.e. random motions.
This is rather unexpected for such gas-rich galaxies, but in line with
the merger interpretation.  Moreover, a few galaxies show peculiar
rotation curves that initially rise rapidly, followed by a Keplerian
(or even faster) decline, strengthening the interpretation that these
galaxies are not in dynamical equilibrium ({\"O}stlin et al. \cite{ostlin01}).

It is, however, well known that the ISM may be subject to other
types of motions than purely gravitational.  Winds and
outflows, for example, are expected to be present in actively star forming
galaxies like these.  Possibly, the \ha \ velocity field traces
feedback rather than the gravitational potential.  Even if
gravitational motions are likely to dominate bulk motions on
the scale of a whole galaxy, this possibility underlines the
need for information on the kinematical properties of
the {\em stellar} component.

The near-IR [Ca{\sc ii}]$_{\lambda\lambda 8498,8542,8662}$ triplet
(hereafter Ca triplet) is well-suited for dynamical studies in
galaxies dominated by young stellar populations (Dressler
\cite{dressler84}). Even in a pure starburst, red supergiants will
in most places contribute to an observable Ca triplet strength, and
its use  in probing complex galaxy kinematics has
been demonstrated in many papers (e.g. Prada et al.
\cite{prada96}; Garcia-Lorenzo et al. \cite{garcia97}; Kormendy
and Bender \cite{kormendy&bender}; M{\'a}rquez et al. \cite{marquez}). 
We have therefore initiated a
programme aimed at studying the stellar kinematics of BCGs, in
particular those with peculiar \ha \ kinematics, by using the
Ca triplet. Here, we are interested in studying the
stellar kinematics as a function of radius, meaning that levels
fainter than the night sky background need to be reached. This is
a challenging task due to the many  sky lines and strong
fringing of most CCDs in this wavelength region. However, the 
red-optimised FORS2 spectrograph at the ESO VLT is ideally suited for
this work.

Kobulnicky and Gebhardt (\cite{kobulnicky&gebhardt}) studied the integrated
dynamics of late type galaxies, among them a few BCGs, by looking at the
[O{\sc ii}]$_{\lambda 3727}$ and Ca{\sc ii} H and K lines.
They found that stars and gas tend to have similar global
profiles, but their results for low mass galaxies were inconclusive.
Moreover, no LBCG was included in their study.
The  Ca H and K lines are not  useful in our case due
to strong emission lines in the blue spectral region.
%Blue compact galaxies have smaller mass than typical $L^\star$ galaxies.
%According to the current hierarchical paradigm, small galaxies form first,
%and giant galaxies are successively built up mergers of smaller galaxies.
%This process that was very common in the early universe, is still going on
%locally at a slower pace (Le F{\`e}vre et al. \cite{lefevre}).
%Hence, our

Galaxies with masses and properties similar to
local LBCGs are common at higher redshift and make a significant
contribution to the cosmic star formation rate 
(e.g. Guzm{\'a}n et al. 1997, 2003).
Hence, local LBCGs may give insights into the early evolution of
galaxies, and have luminosities high enough to be observable even at high
redshift. However, the study of absorption lines relies on a strong
continuum, and at high redshifts it will be challenging to observe
anything but emission lines. As these are sensitive to dynamical
feedback from star formation, it is an important task to
investigate  how well-coupled gaseous and stellar kinematics are
in local galaxies of the same kind.

In this letter we report on the first results of our study: The
first determination of the stellar kinematics of a BCG through the
Ca triplet. In future papers we shall make a more detailed
analysis and present the results for other galaxies in our sample.

   \begin{figure}
   %\resizebox{\hsize}{!}{\includegraphics{eso400_fc.eps}}
   \resizebox{\hsize}{!}{\includegraphics{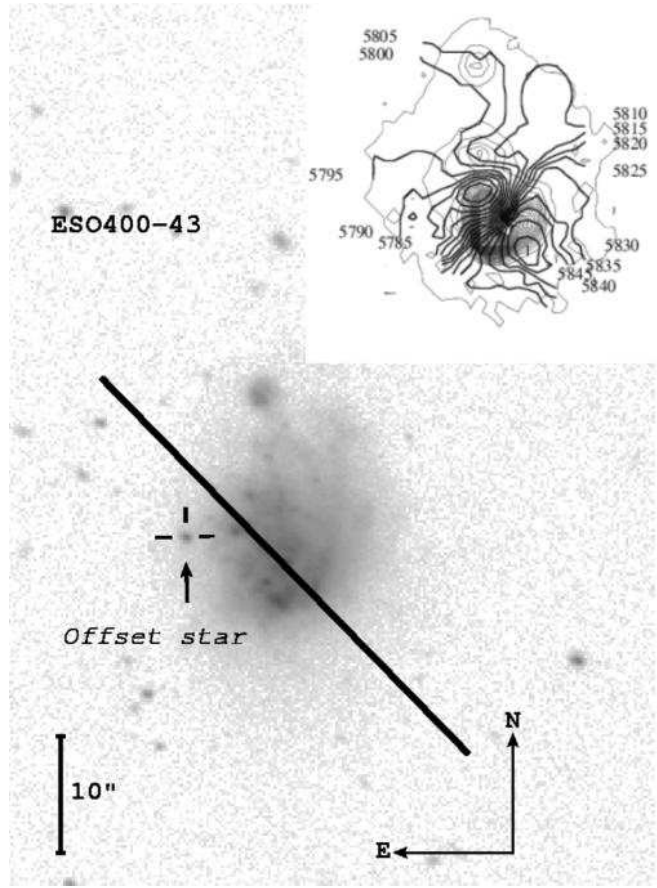}}
   \caption{FORS2 I-band image of ESO\,400-G43 with the slit
    position indicated. The inset in the upper right corner
        shows the \ha \ velocity field from {\"O}stlin et al. (\cite{ostlin99})}
              \label{fc}%
    \end{figure}

%%%%%%%%%%%%%%%%%%%%%%%%%%%%%%%%%%%%%%%%%%%%%%%%%%%%%%%%%%%%%%%%%%%%%%%%%%%%%%%%%%%%

\subsection{ESO\,400-G43}

ESO\,400-G43 is a luminous ($M_B = -19.6$, H$_0 = 75$
\kms${\rm Mpc}^{-1}$) BCG. It was once proposed as a young galaxy
candidate (Bergvall \& J{\"o}rs{\"a}ter \cite{bergvall&jorsater}), but
deep surface photometry has revealed a rather massive
underlying population (Bergvall \& {\"O}stlin
\cite{bergvall&ostlin}). It has a companion galaxy at a projected
distance of 70 kpc, too far for tidal forces to explain the strong
starburst seen in ESO\,400-G43 ({\"O}stlin et al. \cite{ostlin01}).

The \ha \ velocity field is (upper right inset in Fig. \ref{fc}), 
at first sight rather regular, but
after an initial rise to 60 \kms at a radius of 1 kpc, the
rotation curve falls off rapidly, in  contrast to the
photometric mass profile ({\"O}stlin et al. 2001).
%(\cite{ostlin01}).
%({\"O}stlin et al. 2001 \cite{ostlin01}).
The fall is in fact faster
than the Keplerian case, suggesting dynamical disequilibrium. The
apparent rotational energy falls short of the photometric stellar
mass by an order of magnitude. At much larger radii, $\sim 15$
kpc, the H{\sc i} velocity is again close to the central 60 \kms
(Bergvall \& J{\"o}rs{\"a}ter \cite{bergvall&jorsater}) suggesting that a
dynamical instability is responsible for the rapid decline of the optical
rotation curve.

We identify three possible explanations for the apparent discrepancy
between dynamical and photometrical mass.  Gas--star decoupling may be
important, i.e. a stellar velocity field different from the gaseous
one, arising perhaps from outflows triggered by supernova winds or
infall in a merging process.  Alternatively, global gravitational
instabilities might mean that neither stellar nor gaseous motions
trace the potential.  Finally, the low rotational velocities (but not
the shape of the rotation curve) could be explained if velocity
dispersion dominates the gravitational support.

The H$\alpha$ velocity field has a well-defined kinematical centre and
position angle, and was therefore suitable for a study with a single
slit position. Our objective was to derive the stellar rotation curve
and velocity dispersion out to a radius $>$ 2 kpc, making it necessary
to reach surface brightness levels of $\mu_I \ge 21$ mag/arcsec$^2$.

   \begin{figure}
   %\centering
   %\includegraphics{eso400_2d.eps}
   %\sidecaption
   %\includegraphics[width=12cm]{eso400_2d_labels.eps}
   \resizebox{\hsize}{!}{\includegraphics{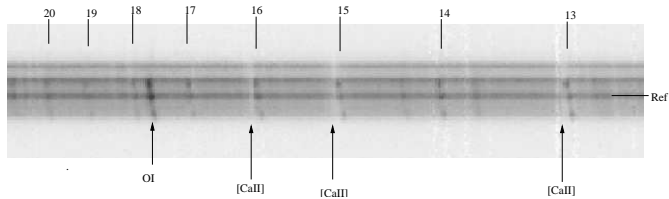}}
   \caption{Two-dimensional spectrum of ESO\,400-G43. The Paschen emission
    lines have been marked and labelled with their number in the Paschen
    series above the spectrum. The Ca{\sc ii} triplet lines and an O{\sc i}
    emission line have been marked below the spectrum.
    On the right, the reference row is indicated. The height of the displayed 
    part of the spectrum is 24\arcsec.}
              \label{f-eso400-2d}%
    \end{figure}

   \begin{figure}
   %\centering
   %\includegraphics[width=15cm]{eso400_pa_sub.eps}
   %\resizebox{\hsize}{!}{\includegraphics{eso400_pa_sub.eps}}
   %\resizebox{\hsize}{!}{\includegraphics{paschen_sub.ps}}
   \resizebox{\hsize}{!}{\includegraphics{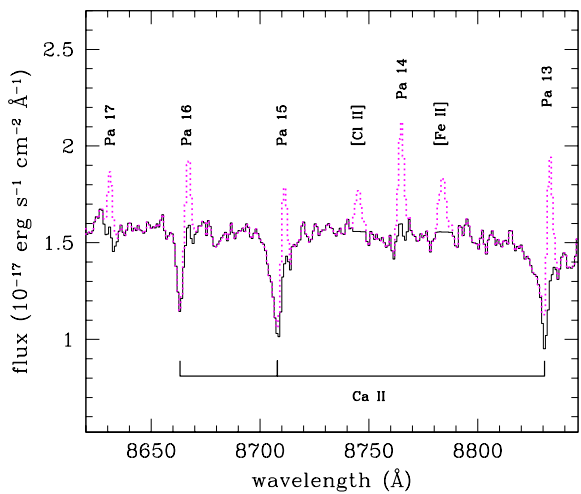}}
      \caption{An example of subtraction of Paschen lines.  In this
    spectrum, the full drawn line shows the subtracted spectrum, which has also
    been interpolated across the emission lines of  [Fe~{\sc ii}] and
    [Cl~{\sc ii}] The dotted line shows the unsubtracted spectrum.
              }
         \label{f-paschen-sub}
   \end{figure}

%%%%%%%%%%%%%%%%%%%%%%%%%%%%%%%%%%%%%%%%%%%%%%%%%%%%%%%%%%%%%%%%%%%%%%%%%%%%%%%%%%%%%%%%%%

\section{Observations and reductions}

The observations of ESO\,400-G43 were obtained between June and August 2003
using the spectrograph FORS2 at the VLT.  We observed one single slit
position with a position angle of $45\degr$ (see Fig. \ref{fc}),
chosen to coincide with the kinematical major axis derived from the
\ha \ velocity field (see Fig. \ref{fc}, and {\"O}stlin et al. 1999). We
used the 1028z grism (covering $\lambda =773$ to $948$ nm) and a slit
width of 0.7\arcsec, resulting in a spectral resolution of $R\sim4000$
near the Ca triplet. The spectra were acquired as a series of 600s
exposures. In between each exposure, the telescope was offset along
the slit.  Using the same setup, six late-type reference stars were
observed, to serve as templates for the cross correlation.

%Our interest was primarily in the spectral region around the Ca triplet. 
We used a somewhat non-standard reduction technique, in
order to ensure effective subtraction of the strong atmospheric 
background emission at these wavelengths:
 	Adjacent pairs of exposures were subtracted from each other, 
each time scaling the subtracted frame so that residuals from the 
OH and O$_2$ lines were minimised. Only those frames resulting a 
in a good sky subtraction were retained for production of the final 
spectrum.
Wavelength calibration was carried out in two dimensions using the 
OH sky lines on the unsubtracted frames.
Each pairwise-subtracted frame was divided by a dome flatfield.
The two-dimensional spectra were shifted to a common origin and  coadded.
Since the offset step along the slit was different in each case, negative 
residuals from objects incidentally falling in the slit cancelled when 
averaging the final frame. The resulting two dimensional spectrum 
(Figure \ref{f-eso400-2d}) was based on 14 frames with a total exposure 
time of 8400 seconds, which was used for the subsequent analysis.
From this final two-dimensional spectrum, one-dimensional extractions
were made, and flux-calibrated 
using spectra of the standard stars LTT 7379 and LTT 7987.  
%No correction was made for telluric absorption features.

   \begin{figure}
   %\centering
   %\includegraphics[width=15cm]{eso400_vel_sig.ps}
   %\resizebox{\hsize}{!}{\includegraphics{plotrot5_2.ps}}
   \resizebox{\hsize}{!}{\includegraphics{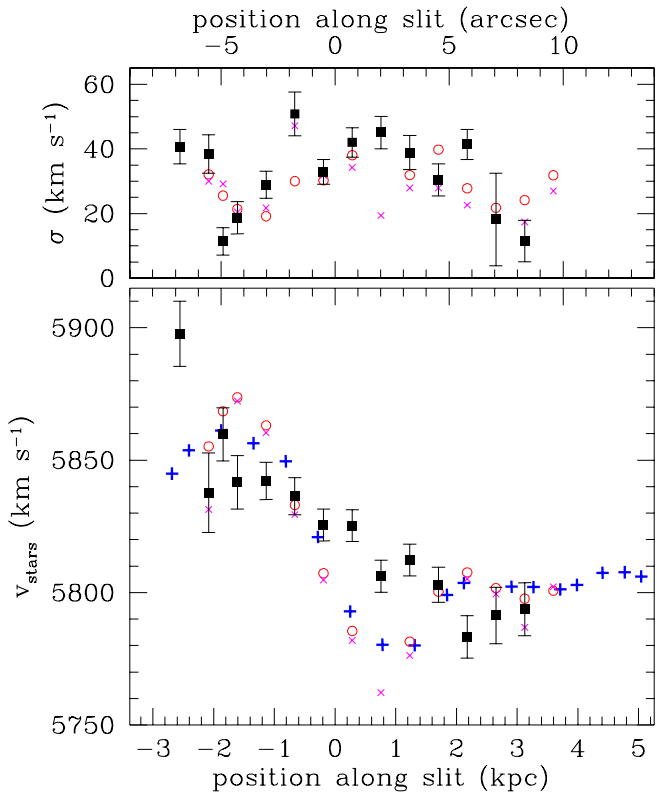}}
      \caption{Variation of heliocentric velocity (bottom) and velocity
	dispersion (top) in ESO\,400-G43, measured from the absorption 
	line spectrum. Squares with error bars represent stellar absorption 
	data. Crosses ($\times$) represent Pa10-lines, and circles  
	[S{\sc iii}] lines. 
	Plus signs (+) show the H$\alpha$ velocities for the same slit 
	angle extracted from Fabry-Perot data ({\"O}stlin et al. 
	\cite{ostlin99}) using a sector opening angle of 5\degr . 
	%Error bars represent the RMS deviation from using different template stars.
	The scale in kpc (bottom) and arcsec (top) is indicated; the zero point
	coincides with the reference row in Fig.
	\ref{f-eso400-2d}.
              }
         \label{f-rotcurve}
   \end{figure}

Our aim was to cross-correlate the extracted spectra with 
reference stars of different types, mainly late-type
giants and supergiants. The galaxy spectra contain also 
narrow emission lines, most of which belong to the H{\sc i}
Paschen series. For each of the Ca triplet lines, there is a
Paschen line (line number 13, 15 and 16 in the series) with only
3--4.5 \AA~ higher rest wavelength. As the Ca absorption features
have intrinsic widths of this order, the Paschen lines had to be
removed before the Ca lines could be used to probe the stellar
velocities.

To achieve this,  we first measured the fluxes and velocities of
the strongest unblended Paschen lines from our spectra, and
compared with the predicted strengths for Case B recombination at
a temperature of 10$^4$ K and density 100 cm$^{-3}$
(Osterbrock 1989, Hummer \& Storey \cite{hummer}).  
Within the  observational uncertainties,
the lines follow the predicted values for zero reddening,
consistent with the low extinction reported by Bergvall \&
{\"O}stlin (\cite{bergvall&ostlin}).
We then modelled the Paschen series from Pa~9 to
Pa~20 for each extracted spectrum, scaled to match the strongest clean
Paschen lines  and subtracted the modelled emission
spectrum (Figure \ref{f-paschen-sub}).  The redshifts and widths of the
modelled lines were determined from the 
strongest Paschen lines in the spectrum.

Finally, we corrected for remaining strong emission lines like O~{\sc i} $\lambda$8446,
[Cl~{\sc ii}] $\lambda$8579 and [Fe~{\sc ii}] $\lambda$8617, bad
pixels and areas of poor sky subtraction by  linearly
interpolating across them.  The spectra, now free of emission lines, were
cross-correlated with the reference stars, using the {\sc iraf} task  
{\tt fxcor} (see Tonry \& Davis \cite{TD}).
In addition to the Ca triplet, these spectra
also contain other weak stellar absorption features which we also
included in the cross correlation.

We calibrated the determination of $\sigma_\star$, the line of
sight stellar velocity dispersion, by broadening the reference star
spectra with Gaussians of various widths, then measuring the
resulting FWHM of the cross-correlation peak. 
%Details will be given in a future paper.  
%Since the
%cross-correlation functions rarely had unambiguously Gaussian
%shapes, we used four different methods of fitting the peak.
%: A
%(base of fitted Gaussian at zero, fitting as much of the central
%peak as is Gaussian in shape), B (base at zero, fitting only the
%central 5-7 velocity bins), C (base at the nearest local minimum
%to the peak's centre) and D (base set assuming the peak resembles
%a sinc function superimposed on a broad Gaussian base).
%This gave us different measures of the velocity and $\sigma_\star$
%for each combination of standard star and extraction along the
%slit.
%In general methods A and B gave the most convincing results, but
%Differences between the methods are very small indicating that our
%procedure is robust. The results are presented in Figure
%\ref{f-rotcurve}.

%______________________________________________________________

\section{Results and discussion}

In Figure \ref{f-rotcurve} we show (filled squares) the stellar
velocities and velocity dispersions along the slit, and the
corresponding data for ionised gas from the Pa10 line (crosses) and
[S{\sc iii}]$_{\lambda 9069}$ (circles). The gas velocity dispersions
have been corrected for the instrumental width.  In addition the
H$\alpha$ velocities extracted from the Fabry-Perot interferometric
data ({\"O}stlin et al. \cite{ostlin99}) are shown (plus signs).  For
clarity, no error bars are plotted for the gas, but uncertainties are
in general much smaller than for the stellar component.

The velocities for the ionised gas agree well with the previous
investigations by Bergvall and J{\"o}rs{\"a}ter
(\cite{bergvall&jorsater}), and {\"O}stlin et al. (\cite{ostlin99}).
The differences between the Pa lines and the H$\alpha$ can be
attributed to the poorer spatial resolution of the latter data. 
The velocities rise to $\pm 55$ \kms (uncorrected for inclination)
over a linear scale of $\pm 1.2$ kpc. Outside
these radii the velocities decline faster than the Keplerian case.
On the approaching side the gas velocities level out at radii $\ge
2$kpc.  However, the difference with respect to the maximum velocity 
is still less than the Keplerian prediction. Hence, the ionised gas 
is not in dynamical equilibrium.

The stellar velocities behave differently.
Firstly, the velocity gradient is much flatter than for the ionised gas.  
At $r=-2.2$ kpc there is a drop in the
rotational velocity (Fig. \ref{f-rotcurve}), but at $r=-2.5$ kpc the rotation 
speed is again high. These two outer points are though the most uncertain 
ones, and there is no strong indication of a decline similar to what is 
seen in the gas. 
On the approaching side, the outer points suggest a modest 
velocity decrease at $r\ge 2.5$ kpc  (Fig. \ref{f-rotcurve}),
where a twist is seen in the \ha \ velocity field (see Fig. 
\ref{fc}, and {\"O}stlin et al. \cite{ostlin99}) as if a spiral arm were present.
Taken together, the stellar velocity field is consistent with solid body 
rotation over the central $\sim$12\arcsec \ (5 kpc) of our 
spectrum, where the S/N is good.

The difference between the minimum and maximum stellar velocity is
$124$ \kms, but this is biased by the uncertain point at $r=-7\arcsec$. 
A straight line fit to the
stellar velocities produces a velocity of $\sim 80$ \kms on 
a scale of 4.5 kpc (12\arcsec). With an assumed inclination of $i=55\degr$
({\"O}stlin et al. \cite{ostlin99}) this implies a rotational
amplitude of $50$ \kms. Using the same simple formula as in
Eq. 1 of {\"O}stlin et al. (\cite{ostlin99}, see also Lequeux 1983)
with $f=0.8$ gives a mass of $1.3\times10^9 M_\odot$, somewhat
lower than the photometric mass inside this radius of
$3^{+3}_{-1}\times10^9 M_\odot$.
The photometric mass profile indicates that the rotation curve
should continue to rise out to a radius of 4 kpc (see Fig. 5 in {\"O}stlin
et al. \cite{ostlin01}). At $r = 2$ kpc  the
predicted circular speed for rotational support is
$80^{+30}_{-15}$ \kms, whereas we measure $40/\sin(i)$ \kms.

This rotational measure assumes that the kinematical position angle
for the stars coincides with that for the gas (45\degr). However, the
major-axis position angle from the broad band isophotes and H{\sc i}
data is closer to 20\degr .  Moreover, the broad-band data suggest a
smaller inclination than 55\degr, while the H{\sc i } kinematics
imply $i\approx 50$\degr (Bergvall and J{\"o}rs{\"a}ter
\cite{bergvall&jorsater}).  H$\alpha$/H$\beta$ mapping ({\"O}stlin \&
Bergvall \cite{ostlin&bergvall}) show that the southeastern side of
the disk is least affected by extinction and thus is probably nearest
to us. 
	With this orientation of the disk, an {\em outflow} would have the 
effect, which we also observe, that the kinematical position angle 
of the gas would be rotated anti-clockwise with respect to  the 
stellar component. 
	Consequently the mass from the stellar rotation
curve may be underestimated, and should be increased by a factor of
1.9 (for $i=55\degr$), or 2.3 if the true inclination is as low as
40\degr. In both cases, the mass inferred from the stellar rotation
would be in fair agreement with the photometric mass.

The velocity dispersion measured from the ionised gas agrees well with
that derived for the stars. Both are however lower than the 49 \kms
derived from Fabry-Perot interferometry ({\"O}stlin et al. 2001). The widths we present in Fig. 4 are, however, close
to the instrumental width and may be underestimated. The agreement
between stars and gas suggests that the widths of the hydrogen
recombination lines reflect motions in response to the gravitational
potential (see Melnick et al. 1987). Under this assumption, {\"O}stlin et
al. (2001) derived a mass of $8\times10^9 M_\odot$. Taking
$\sigma_\star \approx 40$ \kms we estimate the mass supported by
stellar velocity dispersion to be $5\times10^9 M_\odot$.

It is clear that the gaseous velocity field does not reflect dynamical
equilibrium. A model that can reproduce the main features in both the
stellar rotation curve and the H$\alpha$ velocity field reasonably
well is one with solid body rotation with a position angle of 20\degr
and an outflow with a 30 \kms expansion velocity out to a radius of
$\sim 3$ kpc, where the expansion is halted in a shock.  However, this
model is not well constrained.  A two-dimensional stellar velocity
field would be required in order to construct a physical model for the
kinematics and distinguish between the possible explanations given in
Sect. 1.1.

\section{Conclusions}

The gaseous rotation curve of ESO\,400-G43 increases steeply within 1 kpc from
the kinematical centre, then drops faster than the Keplerian
prediction.  This is not seen in the stellar rotation curve which
presents a solid body shape out to $r=2$ kpc.
Hence, stars and gas are dynamically decoupled in this galaxy. 
The reasons for this could be outflows due to feedback from star formation 
or dynamical instabilities in a merging process, and will be discussed 
in future papers.

The stellar and gaseous velocity dispersions agree, but the amount of 
stellar velocity dispersion is somewhat smaller than implied by the photometric 
mass of $12^{+16}_{-4}\times 10^9 M_\odot$ ({\"O}stlin et al. 2001). 
Nevertheless, even allowing for uncertainties in the inclination, the 
stellar velocity dispersion makes a larger contribution than rotation
 to the gravitational support. This is consistent with the picture that the
starburst host galaxy could be of, or will evolve into, elliptical type
(see Bergvall \& {\"O}stlin 2002).

Even though the stellar and gaseous kinematics appear decoupled in this
galaxy, the mass inferred from the line widths of nebular lines and
stellar absorption features agree. It remains to be seen if this is a
general feature of luminous BCGs, but if so, it would lend credibility to
mass estimates of high redshift galaxies that are based on the widths
of narrow nebular lines.

\begin{acknowledgements} 
We thank our colleagues R. Guzm{\'a}n and T. Marquart for stimulating 
discussions and useful suggestions. This work was supported by the 
Swedish Research Council.
\end{acknowledgements}

\end{document}